# Quantum and classical correlations in Bell three and four qubits, related to Hilbert-Schmidt decompositions

Y. Ben-Aryeh

*Physics Department Technion-Israel Institute of Technology, Haifa, 32000, Israel*

E-mail: phr65yb@physics.technion.ac.il

ABSTRACT

The present work studies quantum and classical correlations in 3-qubits and 4-qubits general Bell states, produced by operating with $B_n$ Braid operators on the computational basis of states. The analogies between the general 3-qubits and 4-qubits Bell states and that of 2-qubits Bell states are discussed. The general Bell states are shown to be maximal entangled, i.e., with quantum correlations which are lost by tracing these states over one qubit, remaining only with classical correlations. The Hilbert–Schmidt (HS) decompositions for 3-qubits general Bell states are explored by using 63 parameters. The results are in agreement with concurrence and Peres-Horodecki criterion for the special cases of bipartite system. We show that by tracing 4-qubits general Bell state, which has quantum correlations, over one qubit we get a mixed state with only classical correlations.

## 1. INTRODUCTION

In the present paper we use a special representation for the group $B_n$ introduced by Artin.[1,2] We develop n-qubits states in time by the Braid operators, operating on the pairs $(1,2),(2,3),(3,4)$ etc., consequently in time. Each pair interaction is given by the R matrix satisfying also the Yang-Baxter equation. By performing such unitary interactions on a computational-basis of n-qubits states,[3,4] we will get n-qubits entangled orthonormal general Bell basis of states. The properties of these general Bell states which generalize the properties of the 2-qubits Bell states to n-qubits states ($n$ >2), are analyzed in the present work. The 3-qubits and 4-qubits Bell entangled states are developed explicitly and we show how to extend the calculations for any n-qubits Bell entangled states.

While there are various analyses for quantifying entanglement in bipartite systems,[5–10] these analyses do not apply directly to large n-qubits systems (n>2). In the present work we analyze 3-qubits and 4-qubits systems by using Hilbert-Schmidt (HS) decompositions.[11] For 3-qubits states the analysis is dependent on 63 parameters, while for 4-qubits states it is dependent on 255 parameters! Such analysis is related to the complexity of entanglement for large n-qubits states (n>2), and we show how these parameters can be related to experimental measurements. We find that for 3-qubits general Bell states we have only relatively small number of parameters which are different from zero, while many parameters vanish. We show how the maximal entanglement property of general Bell states is related to the HS parameters.

The present paper is arranged as follows:

In Section 2, I describe the special representation for the Braid operators satisfying $B_n$ Artin group[2] relations, and correspondingly describe the universal R gate, satisfying the Yang-Baxter equation.[12] In Section 3, I develop the use of Braid $B_n$ operators for producing maximal entangled n-qubits states, referred here as general Bell states. We analyze especially the 3-qubits and 4-qubits general Bell entangled states. In Section 4, I show that by tracing 3-qubits Bell state over one qubit, I obtain a mixed state with only classical correlations which is related to concurrence analysis and Peres-Horodecki criterion of bipartite systems. In Section 5 the maximal entanglement property of the general Bell states is related to HS decompositions. I relate the HS decomposition of 3-qubits state to experimental measurements showing which measurements will include quantum correlations and which ones include only classical correlations. I show that by tracing 4-qubits Bell states over one qubit we remain only with classical correlations. In Section 6, I summarize the present results and conclusions.

## 2. The Braid $B_n$ group related to Pauli matrices

The $R$ matrix is given in the present work by[12] :

$$R = \frac{1}{\sqrt{2}} \begin{pmatrix} 1 & 0 & 0 & 1 \\ 0 & 1 & -1 & 0 \\ 0 & 1 & 1 & 0 \\ -1 & 0 & 0 & 1 \end{pmatrix} . \qquad (1)$$

Equation (1) satisfy also a special Yang-Baxter equation given by

$$(R \otimes I) \cdot (I \otimes R) \cdot (R \otimes I) = (I \otimes R) \cdot (R \otimes I) \cdot (I \otimes R) \quad . \tag{2}$$

Here the dot represents ordinary matrix multiplication, the symbol $\otimes$ represents outer product, $I$ is unit $2 \times 2$ matrix, and where $(R \otimes I)$ and $(I \otimes R)$ are matrices of $8 \times 8$ dimensions. The unitary $R$ matrix can be considered in QC as a universal gate,[12] so that all entanglement processes in QC can be described by the $R$ matrix satisfying the Yang Baxter equation, plus the use of local single qubit transformations.

We use for our system the finite representation discovered by Artin,[2] where the generators $\sigma_1, \sigma_2, \cdots, \sigma_{n-1}$ satisfy the $B_n$ group relations:

$$B_n = \left\langle \begin{array}{l} \sigma_1, \sigma_2, \cdots, \sigma_{n-1} \mid \sigma_i \sigma_j = \sigma_j \sigma_i \quad |i-j| > 1 \ ; \\ \sigma_i \sigma_j \sigma_i = \sigma_j \sigma_i \sigma_j \quad |i-j| = 1 \ ; \quad \sigma_i \sigma_i^{-1} = I \end{array} \right\rangle \quad . \tag{3}$$

For the $B_n$ operators $\sigma_1, \sigma_2, \cdots, \sigma_{n-1}$, operating, in our system, on the n-qubits states, we use the representation:

$$\sigma_1 = R \otimes I \otimes I \cdots \otimes I \ ;$$
$$\sigma_2 = I \otimes R \otimes I \cdots \otimes I \ ;$$
$$\sigma_3 = I \otimes I \otimes R \cdots \otimes I \ ;$$
$$\cdot$$
$$\cdot \tag{4}$$
$$\cdot$$
$$\sigma_{n-1} = I \otimes I \otimes I \cdots \otimes I \otimes R$$

Here $R$ is the unitary matrix given by (1), and $I$ is the unit $2 \times 2$ matrix. One should take into account that the operators $R$ and $I$ on the right side of (4) operate on the qubits $1, 2, \cdots, n$, where $R$ entangles two consequent qubits, and the $I$ matrices represent qubits which are non-interacting. One should notice that for 3-qubits system we have in the present Braid group, 2 operators: $\sigma_1 = R \otimes I$ and $\sigma_2 = I \otimes R$ which are described by matrices of $8 \times 8$ dimension. For the 4-qubits system we have 3 operators: $\sigma_1 = R \otimes I \otimes I$, $\sigma_2 = I \otimes R \otimes I$ and $\sigma_3 = I \otimes I \otimes R$, described by matrices of $16 \times 16$ dimension, etc.

The operator $R_{i,i+1}$ entangling $i'th$ and $(i+1)'th$ qubits can be given as:

$$R_{i,i+1} = \frac{1}{\sqrt{2}}(I)_i \otimes (I)_{i+1} + \frac{i}{\sqrt{2}}\Sigma_{x,i} \otimes \Sigma_{y,i+1} \tag{5}$$

Here $\Sigma_{x,i}$ and $\Sigma_{y,i+1}$ are Pauli matrices operating on qubits $i$ and $i+1$, correspondingly. (Notice that we use the notation $\Sigma_x, \Sigma_y, \Sigma_z$ for the Pauli matrices, while the definitions $\sigma_i$ are used for $B_n$ operators). Equation (5) can be described also as

$$R_{i,i+1} = \exp\left[i\frac{\pi}{4}\Sigma_{x,i} \otimes \Sigma_{y,i+1}\right] \tag{6}$$

The exponent in (6) can be related experimentally to Hamiltonian. Eqs. (5-6) are quite important as they relate the present $B_n$ operators to multiplications of Pauli matrices (including the unit matrix $I$).

One should take into account that the operators $R$ and $I$ on the right side of (4) operate in the Braid diagram consequently on the qubits $1, 2, \cdots, n$, where $R$ entangles two consequent qubits and the $I$ matrices represent qubits which are non-interacting.

By using (4) for a 3-qubits system we get the $B_n$ operators

$$\sigma_1 = R \otimes I \quad ; \quad \sigma_2 = I \otimes R \quad . \tag{7}$$

These operators satisfy the Yang Baxter equation:

$$\sigma_1 \sigma_2 \sigma_1 = \sigma_2 \sigma_1 \sigma_2 \quad . \tag{8}$$

I show in the following Sections that that maximal entangled 3-qubits or 4-qubits states can be obtained by operating with $\sigma_1\sigma_2$ or $\sigma_1\sigma_2\sigma_3$ on the 3-qubits or 4-qubits computational basis-states, respectively. Although one can use different multiplications of $\sigma_i$ operators, I found by some calculations that the use of such different multiplications would not lead to maximal entangled states by operating on the computational basis of states.

## 3. Three-qubits and four-qubits Bell-states related to the $B_n$ group

We denote, $|0\rangle = \begin{pmatrix} 1 \\ 0 \end{pmatrix}$ and $|1\rangle = \begin{pmatrix} 0 \\ 1 \end{pmatrix}$ as the two states of each qubit and the subscripts $A, B$ and $C$ will refer to the first, second and third qubit, respectively. The computational basis of 3-qubits states[3,4] is given by the following equivalent forms:

$$|C1\rangle = |0\rangle_A |0\rangle_B |0\rangle_C = \begin{pmatrix}1\\0\end{pmatrix} \otimes \begin{pmatrix}1\\0\end{pmatrix} \otimes \begin{pmatrix}1\\0\end{pmatrix} \quad ; \quad |C2\rangle = |0\rangle_A |0\rangle_B |1\rangle_C = \begin{pmatrix}1\\0\end{pmatrix} \otimes \begin{pmatrix}1\\0\end{pmatrix} \otimes \begin{pmatrix}0\\1\end{pmatrix} \quad ;$$

$$|C3\rangle = |0\rangle_A |1\rangle_B |0\rangle_C = \begin{pmatrix}1\\0\end{pmatrix} \otimes \begin{pmatrix}0\\1\end{pmatrix} \otimes \begin{pmatrix}1\\0\end{pmatrix} = \quad ; \quad |C4\rangle = |0\rangle_A |1\rangle_B |1\rangle_C = \begin{pmatrix}1\\0\end{pmatrix} \otimes \begin{pmatrix}0\\1\end{pmatrix} \otimes \begin{pmatrix}0\\1\end{pmatrix} \quad ;$$

(9)

$$|C5\rangle = |1\rangle_A |0\rangle_B |0\rangle_C = \begin{pmatrix}0\\1\end{pmatrix} \otimes \begin{pmatrix}1\\0\end{pmatrix} \otimes \begin{pmatrix}1\\0\end{pmatrix} \quad ; \quad |C6\rangle = |1\rangle_A |0\rangle_B |1\rangle_C = \begin{pmatrix}0\\1\end{pmatrix} \otimes \begin{pmatrix}1\\0\end{pmatrix} \otimes \begin{pmatrix}0\\1\end{pmatrix} \quad ;$$

$$|C7\rangle = |1\rangle_A |1\rangle_B |0\rangle_C = \begin{pmatrix}0\\1\end{pmatrix} \otimes \begin{pmatrix}0\\1\end{pmatrix} \otimes \begin{pmatrix}1\\0\end{pmatrix} ; \quad |C8\rangle = |1\rangle_A |1\rangle_B |1\rangle_C = \begin{pmatrix}0\\1\end{pmatrix} \otimes \begin{pmatrix}0\\1\end{pmatrix} \otimes \begin{pmatrix}0\\1\end{pmatrix} \quad .$$

The notations $|C1\rangle, |C2\rangle \cdots |C8\rangle$ will refer to the computational basis of states where these states can be described by 8-dimensional vectors. In the $i'th$ entry of the vector $|Ci\rangle$ $(i=1,2,\cdots,8)$ one gets 1, and in all other entries one gets 0. For example,

$$|C1\rangle = \begin{pmatrix}1\\0\\0\\0\\0\\0\\0\\0\end{pmatrix} \quad ; \quad |C3\rangle = \begin{pmatrix}0\\0\\1\\0\\0\\0\\0\\0\end{pmatrix} \quad ; \quad |C7\rangle = \begin{pmatrix}0\\0\\0\\0\\0\\0\\1\\0\end{pmatrix} \quad . \tag{10}$$

The present Bell entangled states for a 3-qubits system are obtained by operating on the states vectors of (9) with $\sigma_1 \cdot \sigma_2 = (R \otimes I) \cdot (I \otimes R)$ . By operating with the matrix $\sigma_1 \cdot \sigma_2$ on the computational basis of states vectors (9) we get the corresponding Bell entangled basis of states for 3-qubits system. The Bell entangled basis of states are then given by superposition of computational states and it is convenient to write them in the following form:

$$2|B1\rangle \equiv |0\rangle_A|0\rangle_B|0\rangle_C - |0\rangle_A|1\rangle_B|1\rangle_C - |1\rangle_A|0\rangle_B|1\rangle_C - |1\rangle_A|1\rangle_B|0\rangle_C \ ;$$
$$2|B2\rangle \equiv |0\rangle_A|0\rangle_B|1\rangle_C + |0\rangle_A|1\rangle_B|0\rangle_C + |1\rangle_A|0\rangle_B|0\rangle_C - |1\rangle_A|1\rangle_B|1\rangle_C \ ;$$
$$2|B3\rangle \equiv -|0\rangle_A|0\rangle_B|1\rangle_C + |0\rangle_A|1\rangle_B|0\rangle_C + |1\rangle_A|0\rangle_B|0\rangle_C + |1\rangle_A|1\rangle_B|1\rangle_C \ ;$$
$$2|B4\rangle \equiv |0\rangle_A|0\rangle_B|0\rangle_C + |0\rangle_A|1\rangle_B|1\rangle_C + |1\rangle_A|0\rangle_B|1\rangle_C - |1\rangle_A|1\rangle_B|0\rangle_C \ ;$$
$$2|B5\rangle \equiv -|0\rangle_A|0\rangle_B|1\rangle_C - |0\rangle_A|1\rangle_B|0\rangle_C + |1\rangle_A|0\rangle_B|0\rangle_C - |1\rangle_A|1\rangle_B|1\rangle_C \ ;$$
$$2|B6\rangle \equiv |0\rangle_A|0\rangle_B|0\rangle_C - |0\rangle_A|1\rangle_B|1\rangle_C + |1\rangle_A|0\rangle_B|1\rangle_C + |1\rangle_A|1\rangle_B|0\rangle_C \ ;$$
$$2|B7\rangle \equiv |0\rangle_A|0\rangle_B|0\rangle_C + |0\rangle_A|1\rangle_B|1\rangle_C - |1\rangle_A|0\rangle_B|1\rangle_C + |1\rangle_A|1\rangle_B|0\rangle_C \ ;$$
$$2|B8\rangle \equiv |0\rangle_A|0\rangle_B|1\rangle_C - |0\rangle_A|1\rangle_B|0\rangle_C + |1\rangle_A|0\rangle_B|0\rangle_C + |1\rangle_A|1\rangle_B|1\rangle_C \ .$$

(11)

Using (4-5), $\sigma_1 \cdot \sigma_2$ can be written as

$$\sigma_1 \cdot \sigma_2 = \frac{1}{2}\left\{\left[(I)_A \otimes (I)_B + i(\Sigma_x)_A \otimes (\Sigma_y)_B\right] \otimes (I)_C\right\} \cdot \left\{(I)_A \otimes \left[(I)_B \otimes (I)_C + i(\Sigma_x)_B \otimes (\Sigma_y)_C\right]\right\} . \quad (12)$$

By using multiplications of Pauli matrices (12) can be simplified into the form

$$\sigma_1 \cdot \sigma_2 = \frac{1}{2}\left\{\begin{array}{l}(I)_A \otimes (I)_B \otimes (I)_C + i(\Sigma_x)_A \otimes (\Sigma_y)_B \otimes (I)_C + \\ i(I)_A \otimes (\Sigma_x)_B \otimes (\Sigma_y)_C + i(\Sigma_x)_A \otimes (\Sigma_z)_B \otimes (\Sigma_y)_C\end{array}\right\} \quad (13)$$

The results obtained in (11) can be related to the form of $\sigma_1 \cdot \sigma_2$ given in (13), as by operating the sum of terms of (13) on the computational state $|Ci\rangle$, we get a corresponding maximal entangled state $|Bi\rangle$ (i =1, 2,…,8). Any one of the 3-qubits Bell states includes superposition of 4 multiplications with equal probability (i.e., real amplitudes are either 1 or -1), and each state in the same multiplication belongs to a different qubit. This property is analogous to the 2-qubit Bell states property, where we have two multiplications with equal probability and the first and second state in each Bell state belong to the first and second qubit, respectively.

The computational basis for 4-qubits states is given by:

$$|C'1\rangle = |0\rangle_A|0\rangle_B|0\rangle_C|0\rangle_D \quad ; \quad |C'2\rangle = |0\rangle_A|0\rangle_B|0\rangle_C|1\rangle_D \quad ;$$
$$|C'3\rangle = |0\rangle_A|0\rangle_B|1\rangle_C|0\rangle_D \quad ; \quad |C'4\rangle = |0\rangle_A|0\rangle_B|1\rangle_C|1\rangle_D \quad ;$$
$$|C'5\rangle = |0\rangle_A|1\rangle_B|0\rangle_C|0\rangle_D \quad ; \quad |C'6\rangle = |0\rangle_A|1\rangle_B|0\rangle_C|1\rangle_D \quad ;$$
$$|C'7\rangle = |0\rangle_A|1\rangle_B|1\rangle_C|0\rangle_D \quad ; \quad |C'8\rangle = |0\rangle_A|1\rangle_B|1\rangle_C|1\rangle_D \quad ;$$
$$|C'9\rangle = |1\rangle_A|0\rangle_B|0\rangle_C|0\rangle_D \quad ; \quad |C'10\rangle = |1\rangle_A|0\rangle_B|0\rangle_C|1\rangle_D \quad ; \qquad (14)$$
$$|C'11\rangle = |1\rangle_A|0\rangle_B|1\rangle_C|0\rangle_D \quad ; \quad |C'12\rangle = |1\rangle_A|0\rangle_B|1\rangle_C|1\rangle_D \quad ;$$
$$|C'13\rangle = |1\rangle_A|1\rangle_B|0\rangle_C|0\rangle_D \quad ; \quad |C'14\rangle = |1\rangle_A|1\rangle_B|0\rangle_C|1\rangle_D \quad ;$$
$$|C'15\rangle = |1\rangle_A|1\rangle_B|1\rangle_C|0\rangle_D \quad ; \quad |C'10\rangle = |1\rangle_A|1\rangle_B|1\rangle_C|1\rangle_D \quad ;$$

These states can be described by 16-dimensional vectors $|C'i\rangle$ where in the i'th entry one gets 1, and in all other entries one gets zero.

Any 4-qubits Bell state $|B'i\rangle$ is obtained by operating with $\sigma_1 \cdot \sigma_2 \cdot \sigma_3 = (R \otimes I \otimes I) \cdot (I \otimes R \otimes I) \cdot (I \otimes I \otimes R)$ on corresponding state $|C'i\rangle$ from the 16 computational vector states (i=1, 2,…,16). It is straightforward to find the $16 \times 16$ matrix $\sigma_1 \cdot \sigma_2 \cdot \sigma_3$ and the complete orthonormal basis of 4-qubits Bell vector-states $|B'i\rangle$, using the present method. For the simplicity of presentation I give here only the 4-qubits state $|B'1\rangle$ which is obtained by operating with $\sigma_1 \cdot \sigma_2 \cdot \sigma_3$ on the computational 16-dimensional vector $|C'1\rangle$, which has 1 in the first entry, and zero in all other entries. I get:

$$\sqrt{8}|B'1\rangle = |0\rangle_A|0\rangle_B|0\rangle_C|0\rangle_D - |0\rangle_A|0\rangle_B|1\rangle_C|1\rangle_D - |0\rangle_A|1\rangle_B|0\rangle_C|1\rangle_D - |0\rangle_A|1\rangle_B|1\rangle_C|0\rangle_D$$
$$- |1\rangle_A|0\rangle_B|0\rangle_C|1\rangle_D - |1\rangle_A|0\rangle_B|1\rangle_C|0\rangle_D - |1\rangle_A|1\rangle_B|0\rangle_C|0\rangle_D + |1\rangle_A|1\rangle_B|1\rangle_C|1\rangle_D \qquad (15)$$

Any Bell 4-qubits state will include superposition of $2^{n-1} = 8$ multiplications with equal probabilities, in which each state in the same multiplication belongs to a different qubit, from the 4 different qubits, as demonstrated in (15) for $|B'1\rangle$. It is quite straight forward to relate these properties to generalization of (13) to $\sigma_1 \cdot \sigma_2 \cdot \sigma_3$, obtaining, by (6), superposition of 8 multiplications of the Pauli matrices, where by operating with each multiplication on the computational vectors $|C'i\rangle$, we get a corresponding term in the maximal entangled Bell state $|B'i\rangle$ (i=1, 2 , …,16). It is easier, however, to operate directly the 16-dimensional matrix $\sigma_1 \cdot \sigma_2 \cdot \sigma_3$ on the computational basis of states.

The method analyzed above can be generalized for implementing any n-qubits Bell states by operating with the matrix $\sigma_1 \cdot \sigma_2 \cdots \sigma_{n-1}$ on the computational states $|C1\rangle, |C2\rangle, \cdots, |C2^n\rangle$. Any Bell state obtained by this method will include $2^{n-1}$ multiplications, with equal probability, where different states in the same multiplication represent different qubits. In our previous examples we found for the 2-qubits -Bell-states $2^{n-1} = 2$ multiplications, for the 3-qubits, $2^{n-1} = 4$ multiplications, and for the 4-qubits state of (15), $2^{n-1} = 8$ multiplications. The application of this method for any n-qubits system is straight forward but the computation becomes quite complicated for a system with a large number of qubits due to entanglement complexity.

**4. Three-qubits Bell entangled states related to concurrence and separability**

Concurrence and separability properties have been analyzed mainly for bipartite systems.[5–10] We are interested, however, in the analysis for 3-qubits and 4-qubits systems. Therefore, we will relate in the next Section, such systems to HS decompositions which leads to a more complete description for entanglement of such states. I would like to show, however, in the present Section that the present analysis for 3-qubits Bell states is in agreement with concurrence and separability properties for bipartite systems.

Considering a pure state $|\Psi\rangle$ of qubits pair, then, the concurrence $C(|\Psi\rangle)$ of this state is defined to be [5–8]

$$C(|\Psi\rangle) = |\langle \Psi | \tilde{\Psi} \rangle| \quad . \tag{16}$$

Here the tilde denotes spin flip operation

$$|\tilde{\Psi}\rangle = \Sigma_y \otimes \Sigma_y |\Psi^*\rangle \quad . \tag{17}$$

$|\Psi^*\rangle$ is given by the complex conjugate of $|\Psi\rangle$, in the standard basis $\{|00\rangle, |01\rangle, |10\rangle, |11\rangle\}$, and $\Sigma_y$ is the Pauli spin operator $\begin{pmatrix} 0 & -i \\ i & 0 \end{pmatrix}$. The spin flip operation takes the state of each qubit in a pure product state to the orthogonal state so that the concurrence of a pure product

state is zero. On the other hand a completely entangled state such as 2-qubits Bell state is invariant under spin flip so that for such state the concurrence $C(|\Psi\rangle)$ takes the value 1 which is the maximal possible value of $C$.

For the more general case in which the density matrix can be mixed, the "spin flip" density matrix is defined to be [5-8]

$$\tilde{\rho}_{A,B} = (\Sigma_y \otimes \Sigma_y)\rho^*_{A,B}(\Sigma_y \otimes \Sigma_y) \qquad (18)$$

Here the asterisk denotes complex conjugation. The product $\rho_{A,B}\tilde{\rho}_{A,B}$ has only real and non-negative values and the square roots of these eigenvalues in decreasing order are $\lambda_1, \lambda_2, \lambda_3, \lambda_4$. Then, the concurrence of the density matrix $\rho_{A,B}$ is defined as

$$C_{AB} = \max\{\lambda_1 - \lambda_2 - \lambda_3 - \lambda_4, 0\} \qquad (19)$$

We find the interesting point that by taking the density matrix of any 3-qubits Bell $|Bi\rangle\langle Bi|$ ($i=1,2,\cdots,8$) entangled state and by tracing over any single qubit (A, or B, or C) the mixed 2-qubits states is obtained with concurrence equal to zero. It quite easy to verify this result by making the calculation for a typical example given as

$$4\rho_{A,B} = 4Tr_C|B1\rangle\langle B1| = (|0\rangle_A|0\rangle_B - |1\rangle_A|1\rangle_B)(\langle 0|_A\langle 0|_B - \langle 1|_A\langle 1|_B)$$
$$+ (|0\rangle_A|1\rangle_B + |1\rangle_A|0\rangle_B)(\langle 1|_A\langle 0|_B + \langle 0|_A\langle 1|_B) \qquad (20)$$

In the standard basis this density matrix can be written as

$$4\rho_{A,B} = \begin{pmatrix} 1 & 0 & 0 & -1 \\ 0 & 1 & 1 & 0 \\ 0 & 1 & 1 & 0 \\ -1 & 0 & 0 & 1 \end{pmatrix} = 4\rho^*_{A,B} \qquad (21)$$

The product $\rho_{A,B}\tilde{\rho}_{A,B}$ is calculated and given by

$$\rho_{A,B}\tilde{\rho}_{A,B} = \rho_{A,B}(\Sigma_y \otimes \Sigma_y)\rho^*_{A,B}(\Sigma_y \otimes \Sigma_y) = \frac{1}{8}\begin{pmatrix} 1 & 0 & 0 & -1 \\ 0 & 1 & 1 & 0 \\ 0 & 1 & 1 & 0 \\ -1 & 0 & 0 & 1 \end{pmatrix} \qquad (22)$$

It is quite easy to find that the square roots of the eigenvalues of (22) are given by $\lambda_1 = \lambda_2 = \frac{1}{2}$ ; $\lambda_3 = \lambda_4 = 0$ so that according to (19) the concurrence is $0$, which means that after tracing over the qubit $C$, we remain with the 2-qubits mixed state which has only classical correlations. The same conclusion is obtained using Peres-Horodecki criterion.[9,10] Following this criterion we partly transpose (PT) the density matrix $\rho_{A,B}$ for the qubit B as $|k\rangle_B \langle l|_B \rightarrow |l\rangle_B \langle k|_B$ ; $l,k = 1,2$ and leave the qubit A unchanged. Then the PT of $\rho_{A,B}$ is given in the standard basis by

$$4\rho_{A,B}(PT) = \begin{pmatrix} 1 & 0 & 0 & 1 \\ 0 & 1 & -1 & 0 \\ 0 & -1 & 1 & 0 \\ 1 & 0 & 0 & 1 \end{pmatrix}. \qquad (23)$$

The eigenvalues of $\rho_{A,B}(PT)$ are $1/2 ; 1/2 ; 0 ; 0$. As they are non-negative $\rho_{A,B}$ is 'separable' so that it includes only classical correlations.

    Although we have made explicit calculation for one example, due to symmetry properties of the entangled states of (11) we get the following general conclusion: By tracing any qubit from any Bell entangled 3-qubits given by (11) we get a mixed state which has zero concurrence and separable density matrix, so that it includes only classical correlations. In a pictorial description: Assuming that we have Bell entangled 3-qubits state and we send the three qubits to Alice, Bob and Charles, respectively, which are far, each from the other. If we ignore any measurement made by one of them (e.g. by Charles) then the other two (e.g. Alice and Bob) can have only classical correlations between them.

    I find also that by tracing any Bell state for large n-qubits system, over any n-2 qubits of such system we get 2-qubits mixed state which is separable with zero concurrence. This result can be demonstrated by tracing the 4-qubits density matrix of (15) over qubits C, and D. Then the calculation leads to the density matrix (21) which as shown previously is separable with zero concurrence. In this way we can relate large n-qubits state to bipartite system. However, I find that for large n-qubits system (n>2) the analysis by HS decomposition will be better than the previous ones,[5–10] but more complicated. We will show in the next Section that by using

the HS decomposition that the general Bell states for the 3-qubits and 4-qubits are maximal entangled, and entanglement properties can be related to HS parameters.

## 5. Properties of entangled Bell states related to HS decompositions

The mean value of Hermitian operator $O$ in an Hilbert-space $H$, described by density matrix $\rho$, is given by

$$\langle O \rangle = Tr(\rho O) \qquad (24)$$

We can find a set of operators $O$ so that (24) can be solved, uniquely for $\rho$. In order to determine the density operator of n-qubits system, $2^{2n} - 1$ real numbers are required (because $\rho$ is Hermitian, satisfying $Tr\rho = 1$). We need therefore $2^{2n} - 1$ independent observable by which we can determine $\rho$ (and for pure states $tr(\rho^2) = 1$).

For 3-qubits system denoted by A, B and C we use the HS representation of the density matrix:[11]

$$\rho = \frac{1}{8} \left\{ \begin{array}{l} (I)_A \otimes (I)_B \otimes (I)_C + (\vec{r} \cdot \vec{\Sigma})_A \otimes (I)_B \otimes (I)_C + (I)_A \otimes (\vec{s} \cdot \vec{\Sigma})_B \otimes (I)_C + (I)_A \times (I)_B \otimes (\vec{f} \cdot \vec{\Sigma})_C \\ + \sum_{m,n=1}^{3} t_{mn} (I)_A \otimes (\Sigma_m)_B \otimes (\Sigma_n)_C + \sum_{k,l=1}^{3} o_{kl} (\Sigma_k)_A \otimes (I)_B \otimes (\Sigma_l)_C \\ + \sum_{i,j=1}^{3} f_{ij} (\Sigma_i)_A \otimes (\Sigma_j)_B \otimes (I)_C + \sum_{\alpha,\beta,\gamma=1}^{3} G_{\alpha\beta\gamma} (\Sigma_\alpha)_A \otimes (\Sigma_\beta)_B \otimes (\Sigma_\gamma)_C \end{array} \right\}$$

. (25)

Here $I$ denotes the unit $2 \times 2$ matrix, $\otimes$ denotes outer product, the three Pauli matrices are represented by $\vec{\Sigma}$, $\vec{r}, \vec{s}$ and $\vec{p}$ are 3-dimensional parameters vectors, and summations over Pauli matrices are given by $\Sigma_m, \Sigma_n, \Sigma_k, \Sigma_l, \Sigma_i, \Sigma_j, \Sigma_\alpha, \Sigma_\beta, \Sigma_\gamma$ $(m,n,k,l,i,j,\alpha,\beta,\gamma = 1,2,3)$.

We find that a pure 3-qubits entangled state is described by 63 parameters: 9 for $\vec{r}, \vec{s}$, and $\vec{p}$, 27 for $t_{mn}, o_{kl}$, and $f_{ij}$, and 27 for $G_{\alpha\beta\gamma}$. These coefficients can be obtained by tracing the multiplication of $\rho$ by the corresponding term in (25). For example:

$$t_{mn} = Tr\{\rho (I)_A \otimes (\Sigma_m)_B \otimes (\Sigma_n)_C\} \; ; \; G_{\alpha\beta\gamma} = Tr\{\rho (\Sigma_\alpha)_A \otimes (\Sigma_\beta)_B \otimes (\Sigma_\gamma)_C\} \; . \qquad (26)$$

I find that the parameters $\vec{r}, \vec{s}$ and $\vec{p}$ are obtained by measurements in one arm of a measurement device, $t_{mn}, o_{kl}$ and $f_{ij}$, are obtained by measurement in the corresponding two arms of the measuring device, and $G_{\alpha\beta\gamma}$ are obtained by the corresponding measurements in the three arms of the measurement device.

A GHZ entangled state [13,14] can be described as:

$$|\psi\rangle = \frac{|1\rangle_A |1\rangle_B |1\rangle_C + |0\rangle_A |0\rangle_B |0\rangle_C}{\sqrt{2}} \quad . \tag{27}$$

By using the density matrix for this state, and (25) we get by straight forward calculations:[11]

$$G_{122} = G_{212} = G_{221} = -1 \quad ; \quad t_{33} = o_{33} = f_{33} = G_{111} = 1 \quad . \tag{28}$$

All other parameters are equal to zero. I find that the quantum correlations in the GHZ state are included in the $G_{\alpha\beta\gamma}$ parameters which are obtained by measurements on the three arms of a measuring device. Although the HS decomposition of the GHZ state includes the parameters $t_{33}, o_{33}, f_{33}$ obtained by measurements in two arms of the device, they do not include quantum correlations, as they are obtained in different two pairs of arms. As one can notice from (25) the $t_{33}$ parameter is obtained by measurements in B and C arms, the $o_{33}$ parameter is obtained from measurements in A and C arms, and $f_{33}$ from arms A and B. In order to get quantum correlations we need to have interference terms in the same pair of arms and since we do not get such interference, quantum correlations are not included in the two arms measurements.

It is interesting to find that the pure maximal entangled states of (11) have similar properties so that all quantum correlations are included in the $G_{\alpha\beta\gamma}$ parameters, which correspond to measurements in three arms. Let us present the HS decompositions for some of the $B_n$ entangled states of (11):

$$\begin{aligned} &For \quad |B_1\rangle: \quad G_{113} = G_{311} = G_{131} = -1 \quad ; \quad t_{22} = o_{11} = f_{33} = G_{333} = 1 \quad ; \\ &For \quad |B_3\rangle: \quad G_{333} = G_{311} = G_{131} = t_{22} = o_{11} = f_{33} = -1 \quad ; \quad G_{113} = 1 \quad ; \\ &For \quad |B_7\rangle: \quad t_{22} = G_{131} = -1 \quad ; \quad o_{11} = f_{33} = G_{333} = G_{113} = G_{311} = 1 \quad . \end{aligned} \tag{29}$$

I find the interesting point that that the absolute values of these parameters are the same for all pure 3-qubits $B_n$ entangled states, and only some of the signs are different as demonstrated in (29), for three $B_n$ entangled states. I notice here again that the two arms parameters $t_{22}, o_{11}, f_{33}$ are obtained from measurements in different pairs of arms so that they do not include quantum correlations. This conclusion is in agreement with the analysis which have been made in Section 4, where it has been shown that by tracing 3-qubits $B_n$ entangled states over one qubit we get a separable state [9,10] with zero concurrence [5–8] which has only classical correlations. So, in order to get the quantum correlations we need to measure the 3-qubits corresponding here to the three arms of a measuring device.

One might define the 3-qubits Bell states as states of GHZ type, but one should take into account that the 3-qubits states of (11) form a complete orthogonal basis of maximal entangled states. Also these states have been related to the $B_n$ group and their production can be obtained by consequent two qubits entangling processes. Each such two-qubits entangling process can be described by (5), where experimentally the Hamiltonian of such process can easily be related to the exponent in Eq. (6).

It is straightforward to generalize the use of density matrix of 3-qubits states given by (25), to the density matrix of 4 qubits system. However, such decomposition will be very complicated, since it depend on 255 ($2^{2n}-1$) terms, where for 4-qubits Bell state, 16 of them are no-vanishing. The quantum correlations of the 4-qubit Bell state will be included in 4 qubits measurements, in analogy to the result that quantum correlations for 3-qubits Bell state are included in 3-qubits measurements. I use in the following analysis a simple method showing by HS decomposition that the 4-qubits Bell state are maximal entangled. By tracing 4-qubits Bell entangled state over one qubit we get a mixed state and the HS decomposition can be made also to such mixed state. I will show in one example that by tracing 4-qubits Bell state over one qubit we lose all quantum correlations and remain with only classical ones. Due to symmetries included in 4-qubits Bell states, these properties are general for any 4-qubits Bell-state.

By tracing the 4-qubits Bell state of (15) we get:

$$8\rho_{A,B,C} = 8Tr_D |B'1\rangle\langle B'1| =$$
$$\{|0\rangle_A|0\rangle_B|0\rangle_C - |0\rangle_A|1\rangle_B|1\rangle_C - |1\rangle_A|0\rangle_B|1\rangle_C - |1\rangle_A|1\rangle_B|0\rangle_C\} \cdot$$
$$\{\langle 0|_A\langle 0|_B\langle 0|_C - \langle 0|_A\langle 1|_B\langle 1|_C - \langle 1|_A\langle 0|_B\langle 1|_C - \langle 1|_A\langle 1|_B\langle 0|_C\} \quad (30)$$
$$+\{|1\rangle_A|1\rangle_B|1\rangle_C - |0\rangle_A|1\rangle_B|0\rangle_C - |1\rangle_A|0\rangle_B|0\rangle_C - |0\rangle_A|0\rangle_B|1\rangle_C\} \cdot$$
$$\{\langle 1|_A\langle 1|_B\langle 1|_C - \langle 0|_A\langle 1|_B\langle 0|_C - \langle 1|_A\langle 0|_B\langle 0|_C - \langle 0|_A\langle 0|_B\langle 1|_C\}$$

The density matrix is given by

$$8\rho_{A,B,C} = \begin{pmatrix} 1 & 0 & 0 & -1 & 0 & -1 & -1 & 0 \\ 0 & 1 & 1 & 0 & 1 & 0 & 0 & -1 \\ 0 & 1 & 1 & 0 & 1 & 0 & 0 & -1 \\ -1 & 0 & 0 & 1 & 0 & 1 & 1 & 0 \\ 0 & 1 & 1 & 0 & 1 & 0 & 0 & -1 \\ -1 & 0 & 0 & 1 & 0 & 1 & 1 & 0 \\ -1 & 0 & 0 & 1 & 0 & 1 & 1 & 0 \\ 0 & -1 & -1 & 0 & -1 & 0 & 0 & 1 \end{pmatrix}. \quad (31)$$

The mixed state density matrix of (31) can be written as

$$8\rho_{A,B,C} = (I)_A \otimes (I)_B \otimes (I)_C + (I)_A \otimes (\Sigma_y)_B \otimes (\Sigma_y)_C + (\Sigma_y)_A \otimes (\Sigma_y)_B \otimes (I)_C + (\Sigma_y)_A \otimes (I)_B \otimes (\Sigma_y)_C \quad (32)$$

Although the HS decomposition includes correlated $\Sigma_y$ operators in two arms of a measuring device they are in different arms pairs: AB or BC or AC. Since they do not include interference terms in the same arms pair, (32) includes only classical correlations. The conclusion is that by tracing the Bell state (15) over one qubit we lose all quantum correlations. Although we analyzed only one example, due to symmetries of Bell states the conclusion is quite general: by tracing over one qubit of any 4-qubit state, all quantum correlations are lost.

## 6. Summary and discussion

In the present work we have developed large n-qubits (n>2) general Bell entangled states, which have properties analogous to those of the 2-qubits Bell states. The unitary matrix $R$, included in the $B_n$ representation for the $\sigma_i$ operators in (2), can be considered in QC as a special universal gate satisfying also the Yang Baxter equation.[12] We use for our systems a finite representation for the $B_n$ group given by (4) as $\sigma_1, \sigma_2, \cdots, \sigma_{n-1}$. In order to get all possible unitary transformations in QC we need to add single qubits transformations to the present universal gates.[12]

The present method for producing general entangled n-qubits Bell states ($n > 2$) has been analyzed in Section 3. By operating on the computational basis of n-qubits states with the unitary multiplication operator $\sigma_1 \sigma_2 \cdots \sigma_{n-1}$, orthonormal entangled Bell states are produced. The general entangled 3-qubit orthonormal Bell states, given in (11), are produced by operating with $\sigma_1 \sigma_2$ on the 3-qubits computational basis of states. 4-qubit orthonormal Bell basis of states are obtained by operating with the Braid multiplication $\sigma_1 \sigma_2 \sigma_3$ on the 4-qubits computational basis of states, as demonstrated in (15) for one example. The analogies between the 3-qubits and 4-qubits Bells and those of the 2-qubitas states have been discussed.

By tracing entangled 3-qubits Bell states over one qubit we get mixed states which are analyzed by concurrence [5–8] and Peres-Horodecki criterion.[9–10]. We demonstrated in (21-24), calculation in which by tracing over one qubit from a certain 3-qubits entangled Bell state we obtain a mixed state with zero concurrence which is separable, so that it has only classical correlations. Due to the symmetry properties of the states of (11) we find that by tracing over any qubit from any 3-qubits Bell state, we lose all quantum correlations. We generalize such property to any n-qubits Bell states where by tracing over any n-2 qubits we get a mixed state with zero concurrence which is separable.

We analyzed the density matrix of 3-qubits system by using HS decomposition which for such system includes 63 parameters. We find that the quantum correlations for 3-qubits Bell state are included in measurements over the 3-qubits. The correlations between measurements of two qubits, are only classical correlations, as in the HS decomposition they are obtained by different arms combinations so that they do not include interference terms necessary for quantum correlations. We have shown, by using HS decompositions that by tracing the 4-qubit Bell state (15) over one qubit we obtain a mixed state which has only classical correlations. Due to the symmetries included in Bell state this property is general for any 4-qubits Bell state.

Our conclusion is that the general Bell states, obtained in the present work by operating with a certain multiplication of $B_n$ operators on the computational basis of states, are maximal entangled states. Maximal entanglement means according to the present approach that the quantum correlations are obtained by the measurement of all qubits of the system, while measurement of a part of the system gives only classical correlations.